\documentclass[prb,twocolumn,showpacs]{revtex4}
\usepackage{xcolor}
\usepackage{amsmath}
\usepackage{amssymb}
\usepackage{graphicx}
\def\ket#1{\left|#1\right\rangle}

\def\braket#1{\left\langle#1\right\rangle}

\begin{document}

\title{Localization and adiabatic pumping in a generalized Aubry-Andr\'e-Harper  model}

\author{Fangli Liu}
\email{liuf0025@e.ntu.edu.sg}

\author{Somnath Ghosh}

\author{Y.~D.~Chong}
\email{yidong@ntu.edu.sg}

\affiliation{School of Physical and Mathematical Sciences and Centre
  for Disruptive Photonic Technologies, Nanyang Technological
  University, Singapore 637371, Singapore}

\begin{abstract}
  A generalization of the Aubry-Andr\'e-Harper (AAH) model is
  developed, containing a tunable phase shift between on-site and
  off-diagonal modulations.  A localization transition can be induced
  by varying just this phase, keeping all other model parameters
  constant.  The complete localization phase diagram is obtained.
  Unlike the original AAH model, the generalized model can exhibit a
  transition between topologically trivial bandstructures and
  topologically non-trivial bandstructures containing protected
  boundary states.  These boundary states can be pumped across the
  system by adiabatic variations in the phase shift parameter.  The
  model can also be used to demonstrate the phenomenon of adiabatic
  pumping breakdown due to localization.
\end{abstract}

\pacs{61.44.Fw, 42.25.Dd, 73.43.Cd}

\maketitle

\section{Introduction}

The Aubry-Andr\'e-Harper (AAH) model \cite{harper,aubry} is
a workhorse for the study of localization and topological states in
one dimension.  It is described by \cite{aubry}
\begin{equation}
  t \left(\psi_{n+1} + \psi_{n-1}\right) +V_1 \cos(Qn+k) \psi_n= E \psi_n,
  \label{AAo}
\end{equation}
where $\psi_n$ is the wavefunction amplitude at site $n$, $t$ is a
nearest-neighbor hopping, and $V_1$, $Q$, and $k$ are the amplitude,
frequency, and phase of the on-site potential.  The model emerges from
the reduction of a two-dimensional (2D) Quantum Hall (QH) system to a
one-dimensional (1D) chain, with $k$ as the quasi-momentum transverse
to the chain.\cite{harper,hofstadter} When the potential is made
$quasiperiodic$ by setting $Q/2\pi$ to an irrational number, the model
has a localization transition \cite{aubry,hofstadter,rig,simon}: all
bulk eigenstates are extended for $0 < V_1 < 2t$, and localized for
$V_1 > 2t$.  The relationship between quasiperiodicity and
localization has been explored in many subsequent variants of the
model.  Typically, altering the potential modulation leads to starkly
different behaviors; some models exhibit mobility edges
\cite{aubry,econ,kohmoto89,sarma09}, while others lack any
localization transition.\cite{tan} One important variant, which
preserves the critical properties of the original AAH model, involves
incommensurate modulations in the off-diagonal hopping
coefficients.\cite{claro,hatsugai,thouless1983,han} In this case, all
states are localized for $V_1 > 2 \,\textrm{max}(t,V_2)$, where $V_2$
is the amplitude of the off-diagonal modulation.\cite{han}

Recently, the AAH model has attracted renewed attention
\cite{kraus,kraus2,verbin,madsen,lijun,Ganeshan}, due to the
realization that it can be implemented experimentally and used to
study the topological properties of 2D bandstructures.  In a
pioneering paper, Kraus \textit{et al.}~demonstrated that an array of
coupled optical waveguides can be used to realize a quasiperiodic AAH
model with purely off-diagonal couplings, and that it is possible to
implement a ``topological pump'' which adiabatically transfers
boundary states across the array by winding the phase of the coupling
modulations.\cite{kraus} Interestingly, it has been shown that AAH
models with on-site and/or off-diagonal modulation can be regarded as
topologically equivalent to Fibonacci lattices of the same
quasiperiodicity.\cite{kraus2,verbin} Madsen \textit{et al.}~have
pointed out, however, that the boundary states occurring in these 1D
lattices are not limited to the quasiperiodic case; they also appear
in commensurate AAH models, and in both cases they can be explained by
dimensional reduction from a topologically non-trivial 2D
system.\cite{madsen} For example, similar boundary states occur in the
period-3 AAH model \cite{lijun}, and in the period-2 model the
boundary states have Majorana-like characteristics.\cite{Ganeshan}
However, incommensurate and commensurate AAH models do significantly
differ in their localization properties.

This paper describes a generalization of the AAH model containing a
tunable phase difference $\phi$ between the on-site and off-diagonal
modulations.  Previous studies of the AAH model have set $\phi = 0$, a
condition that can be naturally derived from a 2D QH system with a
uniform magnetic field.\cite{claro,hatsugai,thouless1983,han} However,
this is an unnecessary restriction in experimental realizations like
the coupled waveguide arrays described above; in these fabricated
structures, the on-site potential and hopping amplitude can be
independently controlled.

As we shall see, the generalized AAH model has several new and
interesting behaviors which have not previously been explored.
Firstly, in incommensurate lattices, a transition between purely
extended and purely localized bulk states can be induced by varying
$\phi$, keeping the modulation amplitudes fixed.  By contrast, in the
original AAH model, localization can only be induced by varying the
modulation amplitudes.  Using a gauge argument, we are able to deduce
the localization phase diagram for arbitrary $\phi$.  Secondly, the
generalized AAH model can form topologically distinct families.  As
shown in Refs.~\onlinecite{kraus2,verbin}, when AAH models are grouped
by $k$ (the phase common to both on-site and off-diagonal
modulations), every bandgap in the $E$ versus $k$ bandstructure is
topologically non-trivial but \textit{equivalent}, regardless of other
model parameters; topologically trivial bandgaps do not appear.
However, when we use the \textit{relative} phase $\phi$ as the pumping
parameter, both types of bandgap can be observed for different
parameter regimes.  This allows us to propose a scheme for observing
topological ``phase transitions'' using a family of 1D AAH models.
Finally, the model provides a convenient way to demonstrate an
interesting property of topological pumps: the failure of pumping in
the presence of localization, due to the breakdown of
adiabaticity.\cite{Molcanov,Feingold,Jansen,Altshuler}

\section{Model}

The generalized AAH model is described by the tight-binding equation
\begin{multline}
  \left\{ t+V_2 \cos\left[\left(n+\tfrac{1}{2}\right)Q + k\right]\right\} \psi_{n+1}
  \\
  + \left\{ t+V_2 \cos\left[\left(n-\tfrac{1}{2}\right)Q + k\right]\right\}\psi_{n-1} \\
  + V_1\cos(nQ + k + \phi)\psi_n = E \psi_n.
  \label{AAm}
\end{multline}
The parameters $t$, $V_1$, $Q$, and $k$ have the same meanings as in
the original AAH model (\ref{AAo}), and $V_2$ is the amplitude of the
modulation in the off-diagonal
hopping.\cite{claro,hatsugai,thouless1983,han} The on-site and
off-diagonal modulations have the same wavenumber $Q$, but the latter
has an additional phase $\phi$.

Previous studies of the AAH model took $\phi = 0$, motivated by the
derivation of the model from the dimensional reduction of a 2D QH
system.  \cite{claro,hatsugai,thouless1983,han,kraus,kraus2,verbin} If
the 2D system is assumed to have isotropic next-nearest-neighbor
hoppings, and the magnetic vector potential is given by the Landau
gauge $\vec{A} = Qy \hat{x}$ (corresponding to a uniform magnetic
field with $Q/2\pi$ flux quanta per unit cell), the resulting 1D
chains have the same frequency $Q$ and phase $k$ in the diagonal and
off-diagonal modulations (i.e., $\phi = 0$).

\begin{figure}
  \centering\includegraphics[width=0.475\textwidth]{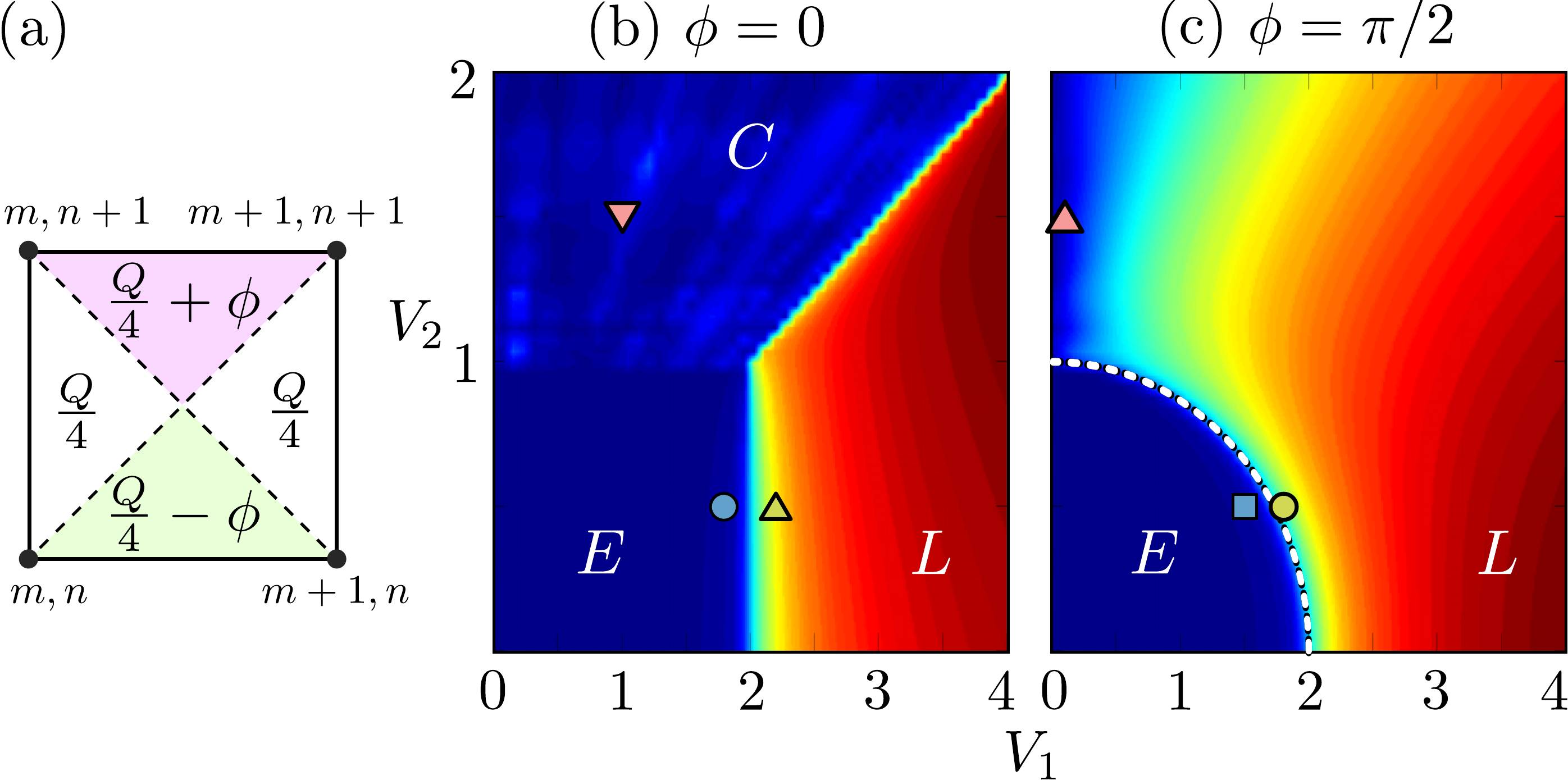}
  \caption{(a) Distribution of magnetic flux in the unit cell which
    gives rise to the generalized AAH model of Eq.~(\ref{AAm}).  (b,c)
    Localization phase diagrams for $\phi = 0$ and $\phi = \pi/2$,
    showing extended ($E$), localized ($L$), and critical ($C$)
    phases.  The heat map shows the ground state's inverse
    participation ratio, with the largest values shown in red.  The
    other parameters are $t = 1$, and $Q = (1+ \sqrt{5})\pi$.  The
    dashed curve in (c) is the theoretical phase boundary, given by
    Eq.~(\ref{phase_boundary}).  The square, circle, and triangle
    symbols indicate the parameter choices for the plots in
    Fig.~\ref{prn}.  }
  \label{fig:phasediag}
\end{figure}

However, the model of Eq.~(\ref{AAm}), with nonzero $\phi$, can also
be generated from a 2D QH system.  The 2D system is simply required to
have nonuniform magnetic flux, with the upper and lower quadrants of
each unit cell receiving extra $\pm \phi$ flux respectively, as shown
in Fig.~\ref{fig:phasediag}(a).  This can be described by a 2D
Hamiltonian consisting of four separate terms:
\begin{align}
  H_1 &= \sum_{mn}\frac{V_1}{2}\,e^{i(nQ+\phi)} \, a_{m+1,n}^\dagger a_{m,n} +
  \mathrm{h.c.} \label{coupling1} \\
  H_2 &= \sum_{mn}t\, a_{m,n+1}^\dagger a_{m,n} + \mathrm{h.c.} \\
  H_3 &= \sum_{mn}\frac{V_2}{2}e^{i(n+\frac{1}{2})Q }
  a_{m+1,n+1}^\dagger a_{m,n} + \mathrm{h.c.} \\
  H_4 &= \sum_{mn}\frac{V_2}{2}\, e^{-i(n+\frac{1}{2})Q}\,
  a_{m,n+1}^\dagger a_{m+1,n} + \mathrm{h.c.} \label{coupling4}
\end{align}
From the phases of the hopping coefficients, one can verify that the
magnetic fluxes are as stated in Fig.~\ref{fig:phasediag}(a).  By
Fourier transforming this Hamiltonian in the $+\hat{x}$ coordinate, we
obtain $H = \sum_k \mathcal{H}(k)$, where $k$ is the quasimomentum in
the $+\hat{x}$ direction and $\mathcal{H}(k)$ is a 1D Hamiltonian
corresponding to the tight-binding equation (\ref{AAm}).

The redistribution of magnetic flux is reminiscent of Haldane's ``zero
field QH'' model, which showed that the topological properties of a QH
system can be altered without changing the net flux per unit
cell.\cite{haldane88} In our model, the flux redistribution described
by $\phi$ affects both the localization and topological properties of
the 1D chains.  Even though a 2D QH system with nonuniform flux may be
challenging to implement, the 1D chains themselves can readily be
realized, as will be discussed in Section \ref{Discussion}.

\section{Localization transition}
\label{Localization transition}

For $\phi = 0$, the localization phase diagram was derived by Thouless
and co-workers \cite{thouless1983,han}, and is shown in
Fig.~\ref{fig:phasediag}(b).  For $V_1 > 2 \,\textrm{max}(t,V_2)$, all
bulk eigenstates are localized; for $V_1, 2V_2 < 2t$, all bulk
eigenstates are extended; and in the rest of the phase space, the
eigenstates are critical.\cite{han,chang97} The localization
transition is driven by the amplitude of the modulations.  In
particular, for $V_2 < t$, localization only depends on $V_1$, and the
critical value is the same as in the purely-diagonal AAH model.

\begin{figure}
  \centering\includegraphics[width=0.47\textwidth]{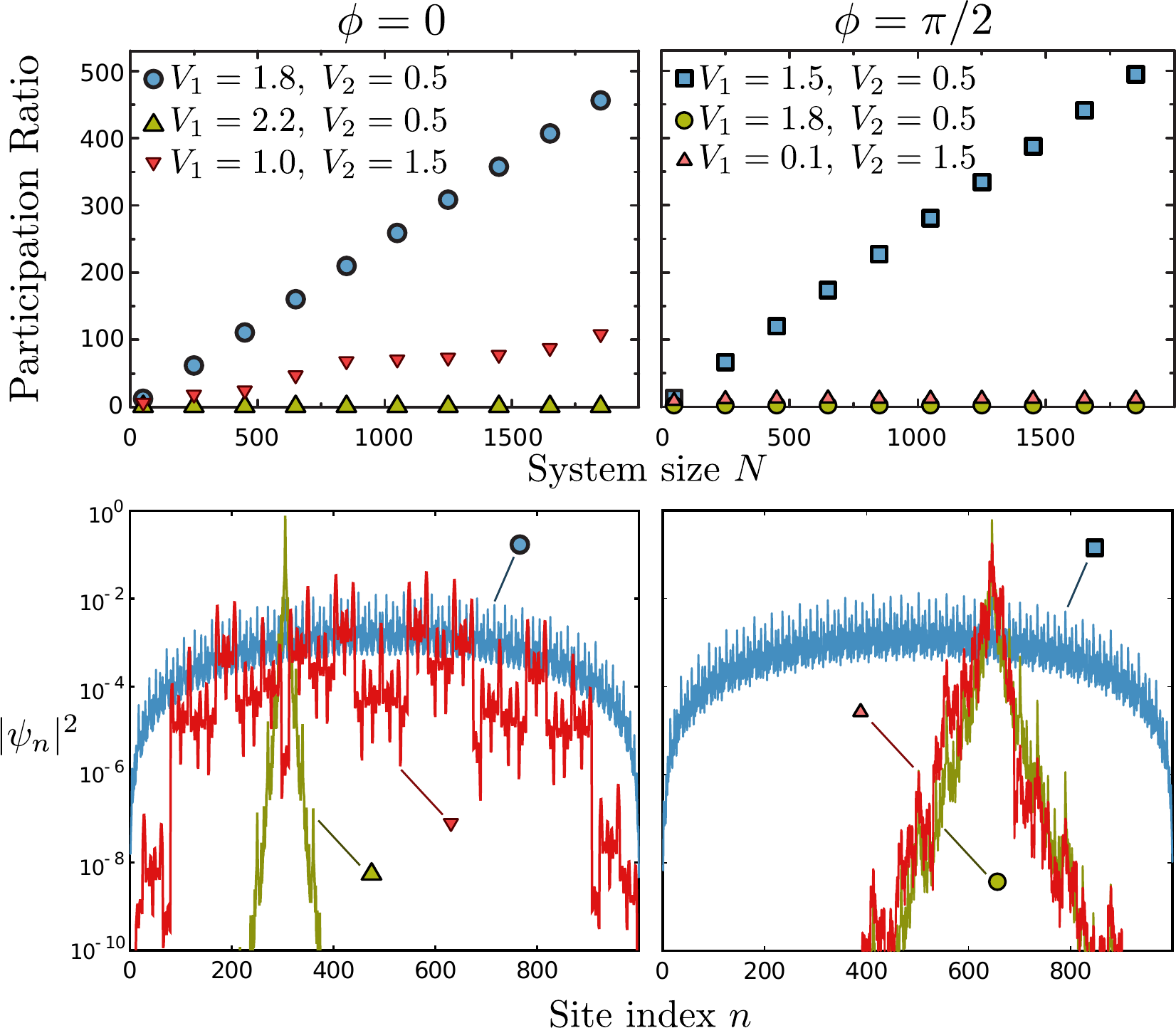}
  \caption{(Color online) Ground state participation ratios and
    probability densities of the generalized AAH model, for $\phi = 0$
    (left) and $\phi = \pi/2$ (right).  The upper plots show the
    scaling of the ground state participation ratio
    $1/\sum_n|\psi_n|^4$ with system size $N$.  The choice of $V_1$
    and $V_2$ parameters is indicated by the matching symbols in the
    phase diagrams in Fig.~\ref{fig:phasediag}; all other parameters
    are the same as in that figure.  The lower plots show $|\psi_n|^2$
    versus site index $n$ for the ground states, for $N = 1000$ and
    the same parameters.  The localized states have participation
    ratios that are constant in $N$, and are located in the localized
    region of the phase diagrams in Fig.~\ref{fig:phasediag}.}
  \label{prn}
\end{figure}

Varying $\phi$ changes the phase diagram.  As shown in
Fig.~\ref{fig:phasediag}(c), for $\phi = \pi/2$ the critical phase
disappears, while the boundary between the extended and localized
phases becomes an arc (which we will derive below).  The heat maps in
Fig.~\ref{fig:phasediag}(b)--(c) show the ground state's inverse
participation ratio (IPR) $\sum_n |\psi_n|^4$, which vanishes for
extended states.\cite{ipr}

Fig.~\ref{prn} shows the ground state participation ratio
($1/\mathrm{IPR}$) and probability density ($|\psi_n|^2$) for
finite-size lattices with different choices of $\phi$, $V_1$ and
$V_2$.  The ground states which are localized are easily identified in
the $|\psi_n|^2$ versus $n$ plots, as well as from the fact that the
participation ratio remains constant with increasing system size $N$.
For the extended states, the participation ratios scale linearly with
$N$, as expected of 1D systems. \cite{ipr} These results are in
agreement with the phase diagrams plotted in
Fig.~\ref{fig:phasediag}(b)--(c).  In particular, observe that for the
parameter choice indicated by the yellow circles (right-hand plots of
Fig.~\ref{prn}), the system is in the localized phase of the $\phi =
\pi/2$ model; by contrast, for the same $V_1$ and $V_2$, the $\phi =
0$ model would be in the extended phase since $V_1 < 2t$.
Furthermore, for the parameter choice indicated by the red
upward-pointing triangles, the system is localized, whereas the $\phi
= 0$ model would be critical.

\begin{figure}
  \centering\includegraphics[width=0.47\textwidth]{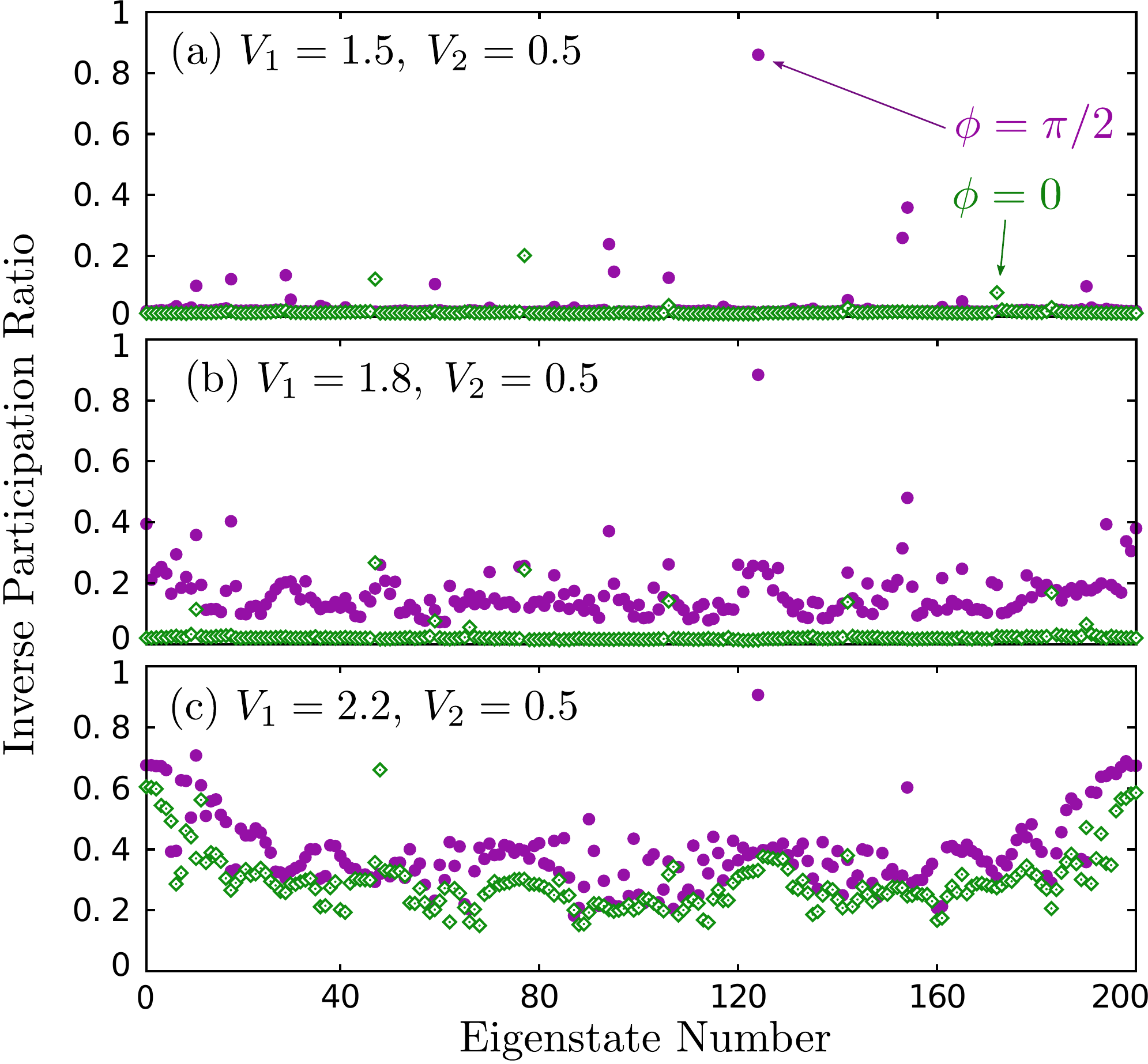}
  \caption{(Color online) Inverse participation ratio (IPR) versus
    eigenvalue number for $V_2 = 0.5$ and (a) $V_1 = 1.5$, (b) $V_1 =
    1.8$, and (c) $V_1 = 2.2$.  Eigenstates are shown for $\phi = 0$
    (green diamonds) and $\phi = \pi/2$ (magenta circles); the system
    size is $N = 200$, and all other parameters are the same as in
    Fig.~\ref{fig:phasediag}.  In (a), bulk states are extended for
    both values of $\phi$; the states with large IPR turn out, upon
    inspection, to be boundary states.  In (b), bulk states are
    extended for $\phi = 0$ and localized for $\phi = \pi/2$.  In (c),
    the states are localized for both values of $\phi$. }
  \label{fig:ips}
\end{figure}

Numerical results show that the excited states have the same
localization properties as the ground states.  Like the original AAH
model, the generalized AAH model lacks a mobility edge.  As
demonstrated in Fig.~\ref{fig:ips}, the bulk eigenstates are either
all extended, or all localized regardless of the eigenstate energy.
(In the finite lattice, however, localized \textit{boundary} states
can occur even in the extended phase, as discussed in the following
sections.)

In the $\phi = 0$ AAH model, the localization behavior has long been
understood to be tied to Aubry-Andr\'e duality: the model is
spectrally invariant under the exchange $t \leftrightarrow V_1/2$,
which maps localized states to extended states and vice versa (and
this remains true when off-diagonal modulations are included).
\cite{aubry,hatsugai,thouless1983,han}

The $\phi \ne 0$ model does \textit{not} obey Aubry-Andr\'e duality.
Still, some of the analytic tractability of the original AAH model
carries over to the generalized model.  We take the QH system
described by Eqs.~(\ref{coupling1})--(\ref{coupling4}), and apply the
gauge transformation
\begin{equation}
  \vec{A}_{mn} \rightarrow \vec{A}_{mn} - nQ \,\hat{x} + mQ\, \hat{y}.
  \label{gauge transform}
\end{equation}
Fourier transforming the 2D Hamiltonian in the $-\hat{y}$ coordinate
then yields a 1D Hamiltonian corresponding to the following
tight-binding equation:
\begin{multline}
  \left\{ \frac{V_1}{2} e^{i \phi}+ V_2 \, \cos\left[\left(m+\tfrac{1}{2}\right)Q + k'\right]\right\} \psi_{m+1}
  \\
  + \left\{ \frac{V_1}{2} e^{-i \phi} + V_2  \cos\left[\left(m-\tfrac{1}{2}\right)Q + k'\right]\right\}\psi_{m-1} \\
  + 2t \cos(mQ + k')\psi_m = E \psi_m.
  \label{AA2}
\end{multline}
Thus, the model is dual under a combination of exchanging $t
\leftrightarrow V_1/2$, and moving the relative phase $\phi$ into the
off-diagonal hopping term.  Note that this reduces to the usual
Aubry-Andr\'e duality for $\phi = 0$.

We can deduce the localization phase boundary with the aid of
Eq.~(\ref{AA2}), together with an argument due to
Thouless.\cite{thouless1983} The Thouless argument provides a lower
bound for the measure of the spectrum; although originally given for
the $\phi = 0$ case, it can be adapted to the model of Eq.~(\ref{AA2})
for $\phi \ne 0$, as shown in Appendix \ref{Localization property}.
Using the principle that the measure of the spectrum vanishes at the
localization transition \cite{aubry,hofstadter}, we find that the
localization phase boundary is described by
\begin{equation}
  \sum_{\pm} \sqrt{\left(V_1/2\right)^2 \pm V_1 V_2 \cos\phi + V_2^2} = 2t.
  \label{tran_eq}
\end{equation}
For $\phi = 0$, Eq.~(\ref{tran_eq}) bounds a rectangular region $V_1
<2t$, $V_2 < t$, which corresponds to the extended phase shown in
Fig.~\ref{fig:phasediag}(b); Aubry-Andr\'e duality then implies the
phase boundary $V_1 = 2 \,\textrm{max}(t,V_2)$, which describes the
localized phase.\cite{han} On the other hand, for $\phi = \pi/2$,
Eq.~(\ref{tran_eq}) reduces to the semi-elliptical arc
\begin{equation}
  \left(V_1/2\right)^2 + V_2^2 = t^2,
  \label{phase_boundary}
\end{equation}
which agrees with the phase diagram shown in
Fig.~\ref{fig:phasediag}(c).

For intermediate values of $\phi$, Eq.~(\ref{tran_eq}) is also in
excellent agreement with numerical results, as shown in
Figs.~\ref{abphase} and \ref{iprv1} in Appendix \ref{Localization
  property}.  Furthermore, we find that the critical phase in the
$\phi = 0$ phase diagram is unstable to variations in $\phi$; for
small $\phi \ne 0$, the entire region outside the extended phase
consists of purely localized states.

\section{Topological properties}
\label{Topological properties}

AAH models can be topologically characterized by noting that a family
of AAH chains with different $k$'s is essentially 2D, with $k$ acting
as an additional compact dimension.  When $Q$ is set to a rational
approximant $2\pi q/p$, the 2D system has well-defined topological
invariants in the form of the Chern numbers of the $p$ bands
\cite{TKNN}.  Because the Chern fluxes are independent of $k$, $V_1$
and $V_2$ for $p \gg 1$, Kraus \textit{et al.}~argued that AAH models
of the same $Q$ can be regarded as being topologically equivalent and
non-trivial.\cite{kraus} In another work, it was shown that the AAH
model can be continuously deformed into a Fibonacci quasicrystal
\cite{kraus2}; this was confirmed by an optical lattice experiment
showing that an AAH lattice and a Fibonacci lattice can be smoothly
connected without closing the bulk gap.\cite{verbin} In this context,
topological transitions are only observed between AAH models with
different modulation frequencies; for fixed $Q$, one cannot induce a
topological transition in a manner similar to the localization
transition, i.e.~by varying the model parameters $t$, $V_1$, or $V_2$.

The generalized AAH model allows for a richer set of topological
behaviors.  If we use the phase $\phi$ as a winding parameter, instead
of $k$, then the AAH model can exhibit \textit{either} topological
trivial \textit{or} non-trivial bandstructures.  We can also induce
topological transitions between these two types of bandstructures by
varying the model parameters.

\begin{figure}
  \centering\includegraphics[width=0.478\textwidth]{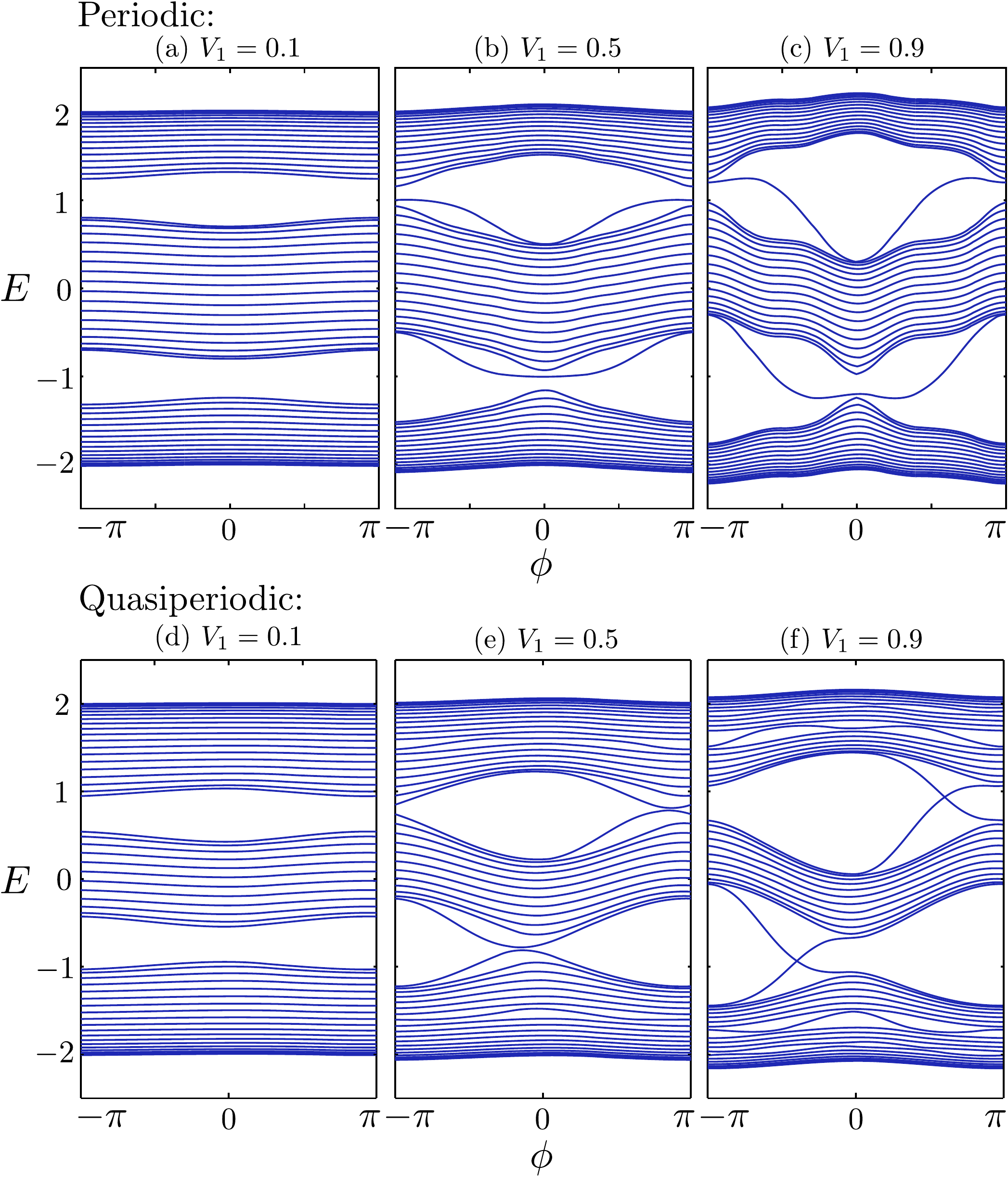}
  \caption{(Color online) Bandstructures of $E$ versus $\phi$ for the
    generalized AAH model.  (a)--(c) Bandstructures of period-3
    lattices ($Q = 10\pi/3$), for $V_1 = 0.1$, $0.5$, and $0.9$.
    (d)--(f) Bandstructures of quasiperiodic lattices ($Q = (1+
    \sqrt{5})\pi$), for those same values of $V_1$.  The other model
    parameters are fixed at $V_2 = 0.25$, $t = 1$, $k = 0$, and $N =
    50$. }
  \label{fig:projected}
\end{figure}

It is important to note, at this point, that quasiperiodicity is not
necessary for studying the topological properties of these 1D model
families.  As previous authors have noted, topological boundary states
can appear in periodic and quasiperiodic systems
alike. \cite{madsen,lijun}

Fig.~\ref{fig:projected} shows the $E$ versus $\phi$ bandstructures
for the generalized AAH model.  For a fixed lattice size $N = 50$, two
sets of results are shown: (i) period-3 lattices with $Q/2\pi = 5/3$,
and (ii) quasiperiodic lattices with $Q/2\pi = (1+\sqrt{5})/2$.  In
both cases, the bandgaps are observed to be free of boundary states
for small $V_1$; as $V_1$ is increased (with $V_2$ and other
parameters fixed), the bulk bandgaps close and then re-open, and for
large $V_1$ the re-opened gaps are spanned by boundary states.  There
is thus a transition from a topologically trivial phase to a
non-trivial phase, for AAH chains of the \textit{same} periodicity.

The topological nature of the boundary states can be demonstrated by
adiabatically varying $\phi$, similar to the adiabatic variation of
$k$ discussed in Ref.~\onlinecite{kraus}.  In the topologically
trivial phase, any boundary states that exist within the bandgaps (due
to finite-size effects) are not topologically protected, and will
remain confined to the same boundary as $\phi$ is varied.  In the
topological phase, however, boundary states that span the bandgaps, as
shown in Fig.~\ref{fig:projected}(c), are topologically protected, and
varying $\phi$ pumps them across the chain.

To understand the topological transition, we study the period-3 model
with $Q/2\pi = 5/3$.  Its Bloch states are the eigenvectors of a
$3\times3$ effective Hamiltonian which depends on $\phi$ and the
quasimomentum along the chain.  We can locate the gap closings, and
within each gapped phase we can calculate the Chern numbers of the
bands, which characterize the topology of the
bandstructure.\cite{TKNN} The details are given in Appendix
\ref{Topological phase transition}.  The relevant topological phase
boundary is found to occur along the curve
\begin{equation}
  V_1 = \frac{V_2}{2} \,\left( \frac{4t - V_2}{t - V_2}\right).
\end{equation}
As $V_1$ increases past the critical value, the bandstructure goes
from topologically trivial, with Chern numbers $\{0,0,0\}$, to
topologically nontrivial, with Chern numbers $\{-1,2,-1\}$.  For $V_1,
V_2 \ll t$, the phase boundary occurs at approximately $V_1 \approx
2V_1$.  This transition of the period-3 model appears to be a good
match for the topological transition of the quasicrystal shown in
Fig.~\ref{fig:projected}(d)--(f).

\begin{figure}
  \centering\includegraphics[width=0.478\textwidth]{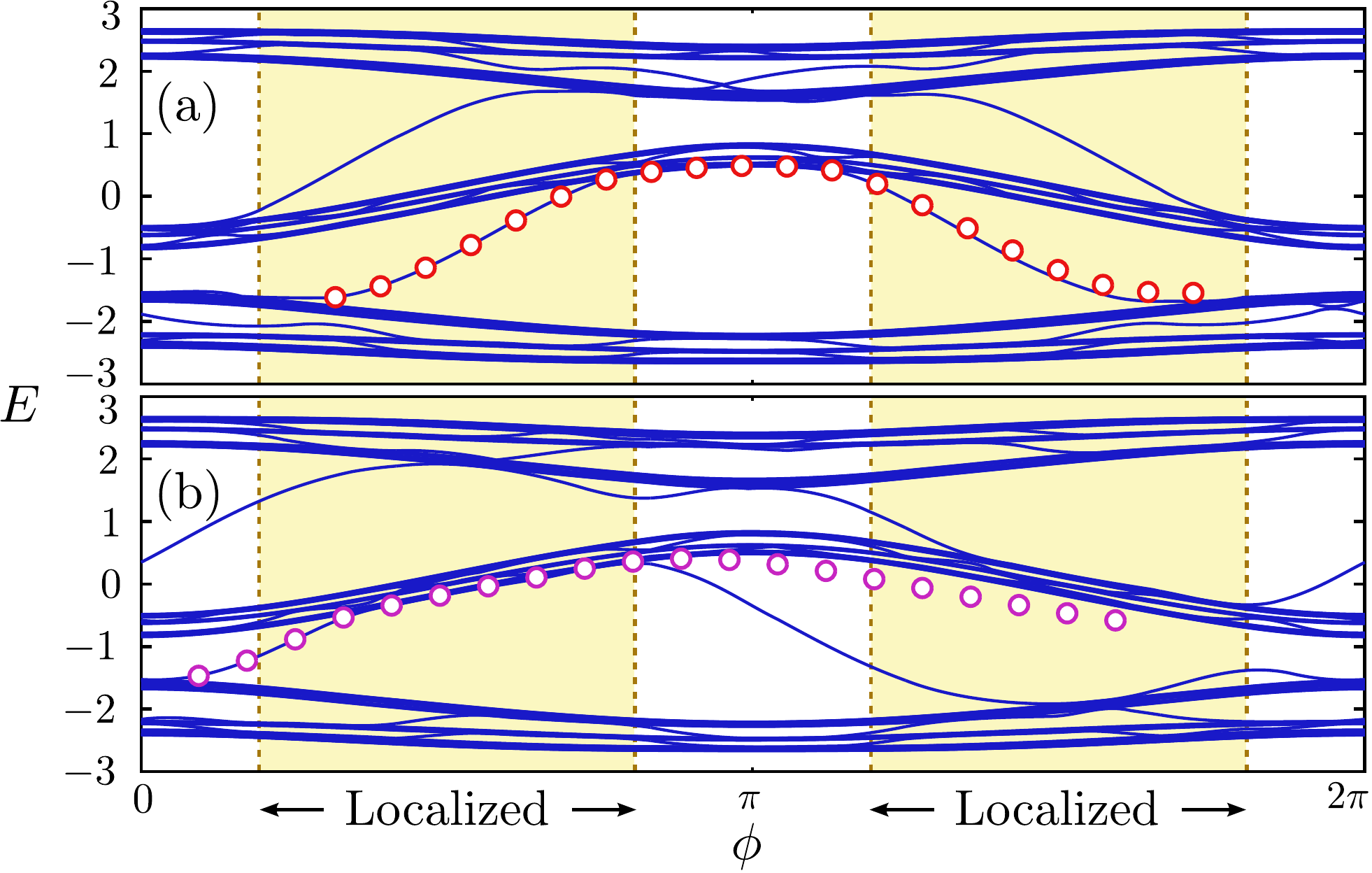}
  \caption{(Color online) Bandstructures of $E$ versus $\phi$ for (a)
    $k = 0$ and (b) $k = 0.3\pi$, with $V_1 = 1.9$, $V_2 = 0.5$, $t =
    1$, $Q = (1+ \sqrt{5})\pi$, and $N = 101$.  Vertical dashes
    indicate the $\phi$ intervals over which the bulk states are
    localized.  The red and purple circles show the expectation value
    $\braket{\psi(t)|E|\psi(t)}$, starting from a boundary state and
    taking $\phi(t) = 10^{-5}t$. }
  \label{fig:endstates}
\end{figure}

\begin{figure}
  \centering\includegraphics[width=0.478\textwidth]{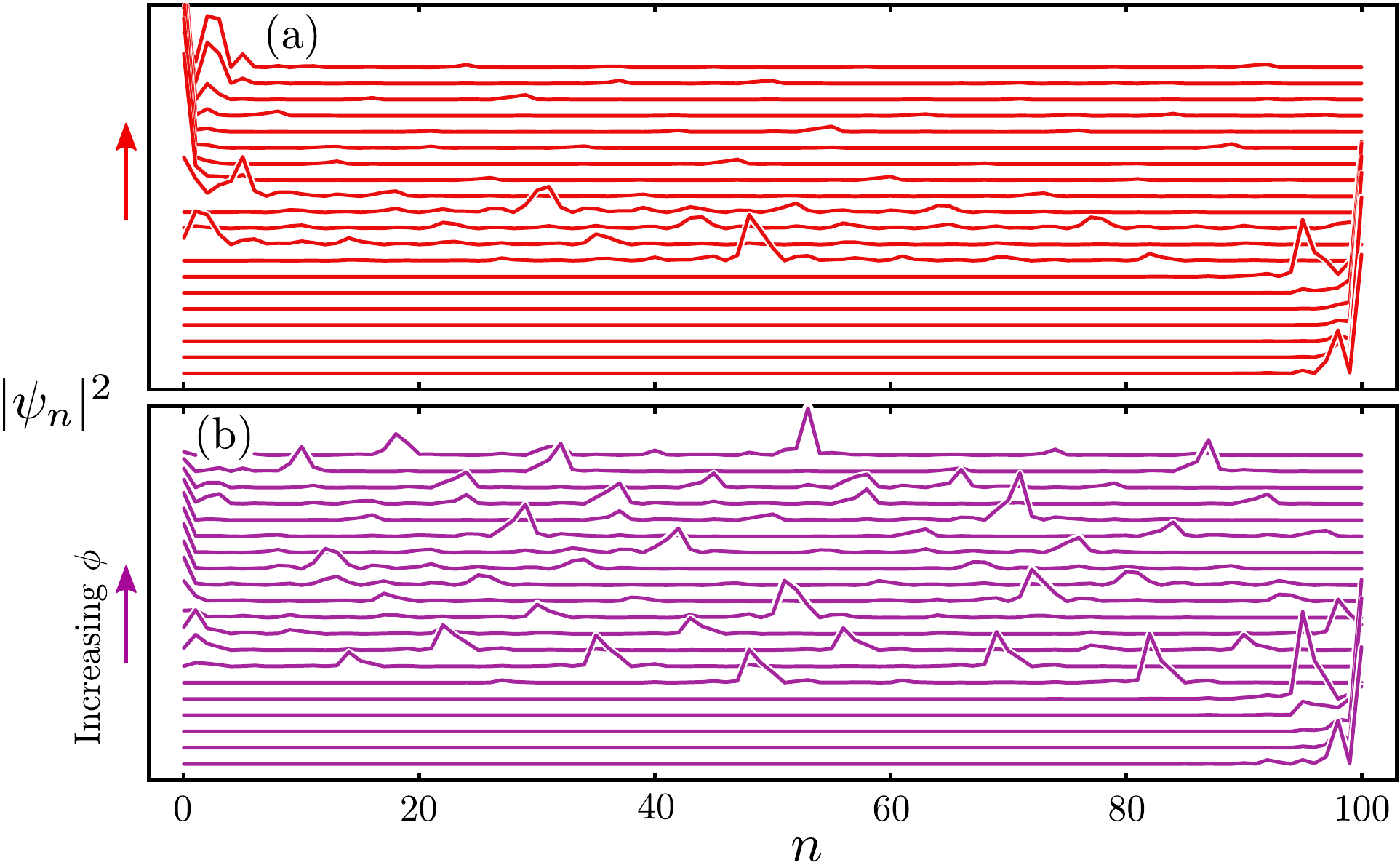}
  \caption{(Color online) Plots of $|\psi_n|^2$ versus chain index $n$
    at subsequent times, based on numerical solutions of the
    time-dependent Schr\"odinger equation with a slowly-varying
    $\phi(\tau) = 10^{-5}\tau$.  The other model parameters are the
    same as in Fig.~\ref{fig:endstates}, with (a) $k = 0$ and (b) $k =
    0.3\pi$.  Each $|\psi_n|^2$ plot corresponds to one of the circles
    in Fig.~\ref{fig:endstates}.  }
  \label{fig:psipump}
\end{figure}

\section{Discussion}
\label{Discussion}

The generalized AAH model is feasible to realize using optical
waveguide lattices \cite{waveguides,waveguides2,waveguides3} or cold
atom systems.\cite{Roati} Such systems have been used to demonstrate
Anderson localization in disordered lattices \cite{Segev07, Lahini08},
localization in AAH chains \cite{Lahini09,Roati}, and adiabatic
pumping of boundary states in off-diagonal AAH chains.\cite{kraus} The
rich physical behavior of the generalized AAH model motivates the
implementation of simultaneous, independently-variable on-site and
off-diagonal modulations in such experiments, so as to be able to tune
the $\phi$ parameter.

In the context of optical waveguide lattices, for instance, the
topological transition described in Section \ref{Topological
  properties} can be demonstrated using an array with two adjacent
regions of the same $Q$ but different $t$, $V_1$, and/or $V_2$.  If
the parameters are chosen so that the two regions are topologically
distinct, then there will be some $\phi$ such that light injected at
the interface is localized (due to overlap with a topological boundary
state) rather than being scattered into the bulk.  If the regions are
topologically equivalent, the existence of boundary states will not be
similarly guaranteed, but will instead depend sensitively on the
parameter choices and interface conditions.

Optical lattices have been used extensively for studying the physics
of localization, and the generalized AAH model provides an unusual
opportunity to examine how localization affects the adiabatic pumping
of topological boundary states.  Pumping involves a boundary state
adiabatically merging into a bulk band and becoming extended, then
evolving into a boundary state at the opposite end.\cite{kraus}
However, this adiabatic process can break down when the bulk states
become localized, due to the suppression of minimum level spacings in
the localized regime.\cite{Molcanov,Feingold,Jansen,Altshuler} This
can be studied in a controlled way using the incommensurate
generalized AAH model, since both localization and pumping are driven
by the $\phi$ parameter.  Fig.~\ref{fig:endstates} shows a situation
in which the bulk states are extended and localized at different
values of $\phi$.  The parameter $k$ changes the dispersion of the
boundary states, though not the bulk bands; hence, we can use $k$ to
control whether a boundary state joins a band in the extended or
localized regime.  We simulate pumping by numerically solving
$i\partial_\tau \ket{\psi(\tau)} = H[\phi(\tau)] \ket{\psi(\tau)}$,
starting from a boundary state and increasing $\phi$ slowly (in the
context of waveguide arrays, $\tau$ is the axial spatial coordinate
\cite{waveguides}).  Fig.~\ref{fig:endstates}(a) and
Fig.~\ref{fig:psipump}(a) show successful pumping of a boundary state.
For a different $k$ (with $d\phi/d\tau$ and all other parameters kept
the same), the boundary state merges into a localized bulk, and this
causes the adiabatic pump to fail as shown in
Fig.~\ref{fig:endstates}(b) and Fig.~\ref{fig:psipump}(b).

In summary, we have shown that a generalization of the AAH model has
far-reaching implications for its localization and topological
properties.  These phenomena should be observable with existing
experimental platforms.  Features remaining to be explored include the
fractal characteristics of the $E$ versus $\phi$ bandstructures, and
the robustness of the boundary states against disorder.

\section{Acknowledgments}

We thank M.~Pasek, J.~C.~Pillay, M.~C.~Rechtsman, H.~L.~Wang,
Y.~Shikano, and Y.~Y.~Ma for their helpful comments.  This research
was supported by the Singapore National Research Foundation under
grant No.~NRFF2012-02, and by the Singapore MOE Academic Research Fund
Tier 3 grant MOE-2011-T3-1-005.

\appendix

\section{Localization phase boundary}
\label{Localization property}

In this appendix, we adapt Thouless' derivation of a bound for the AAH
model's spectral measure\cite{thouless1983} to the generalized AAH
model.  This yields the localization phase boundary (\ref{tran_eq}).
As discussed in Section \ref{Localization transition}, applying the
gauge transformation (\ref{gauge transform}) to the 2D Hamiltonian
given by Eqs.~(\ref{coupling1})--(\ref{coupling4}) gives $H' = H_1' +
\cdots + H_4'$, where
\begin{align}
  H_1' &= \sum_{mn}\frac{V_1}{2} e^{i\phi}\, a_{m+1,n}^\dagger a_{m,n} +
  \mathrm{h.c.} \label{coupling1a} \\
  H_2' &= \sum_{mn}t\, e^{-imQ}\, a_{m,n+1}^\dagger a_{m,n} + \mathrm{h.c.} \\
  H_3' &= \sum_{mn}\frac{V_2}{2}e^{-i(m+\frac{1}{2})Q }
  a_{m+1,n+1}^\dagger a_{m,n} + \mathrm{h.c.} \\
  H_4' &= \sum_{mn}\frac{V_2}{2}\, e^{-i(m+\frac{1}{2})Q}\,
  a_{m,n+1}^\dagger a_{m+1,n} + \mathrm{h.c.} \label{coupling4a}
\end{align}
Fourier transforming in $\hat{y}$ reduces $H'$ to a family of 1D
Hamiltonians $\mathcal{H}'(k')$ corresponding to Eq.~(\ref{AA2}).

We consider $Q = 2\pi q/p$, with $q,p\in \mathbb{Z}$ and $p$ odd
(which we write as $p = 2s+1$), and look for eigenstates of the
infinite 1D chain satisfying
\begin{equation}
  \psi_{m+p} = e^{ikp} \;\psi_{m}.
\end{equation}
(Note that $\psi_{m}$ also depends implicitly on $k'$.)  Due to gauge
symmetry, the spectrum of $\mathcal{H}'(k')$ is independent of the
phase of the hopping amplitudes, so we can replace Eq.~(\ref{AA2})
with the modified tight-binding equation
\begin{multline}
  \mathcal{V}_{m-1,k'}\,\psi_{m-1} + 2t \cos(mQ + k')\psi_m \\ +
  \mathcal{V}_{mk'}\, \psi_{m+1} = E \psi_m,
  \label{BB2}
\end{multline}
where
\begin{equation}
  \mathcal{V}_{mk'} \equiv \left|\frac{V_1 }{2}e^{i\phi} + V_2 
  \cos\left[\left(m+\tfrac{1}{2}\right)Q + k'\right] \right|.
  \label{BB3}
\end{equation}

To find a bound for the spectral measure, we focus on the
high-symmetry points (i) $k = k' = 0$ and (ii) $k = \pi/p$, $k' =
\pi$.  For $k = 0,\pi/p$, the solutions can be split into those that
are symmetric or antisymmetric about the points $m=0$ and $m=s$.  For
the solutions that are symmetric about $m=0$, we can define the
variables $a_0= \sqrt{2} \psi_0$ and $a_m= \psi_m + \psi_{-m}= 2
\psi_m$ for $1 \le m \le s$.  For both the $k' = 0$ and $k' = \pi$
cases, these satisfy:
\begin{widetext}
\begin{align}
  2 t \cos(k') \, a_0 +\sqrt{2} \mathcal{V}_{0k'} \, a_1
  &= E^{+ \pm} \,a_0 \label{e22} \\
  \sqrt{2} \mathcal{V}_{0k'} \, a_0 + 2t \cos(Q+k') \,a_1
  + \mathcal{V}_{1k'} \, a_2 &= E^{+ \pm} a_1 \\
  \mathcal{V}_{m-1,k'} \,a_{m-1}
  + 2 t \cos(m Q+k^{\prime}) \,a_m
  + \mathcal{V}_{mk'}\, a_{m+1}
  &= E^{+ \pm} \,a_m \;\;\;\mathrm{for}\;\; 2\leq m <s \\
  \mathcal{V}_{s-1,k'} \,a_{s-1} + 2t \cos(s Q+k') \,a_s
  \pm \mathcal{V}_{sk'}\, a_s &= E^{+ \pm} \,a_s.
\label{e25}
\end{align}
The $\pm$ signs in these equations denote solutions that are symmetric
about $p/2$ (for $k = 0$) and antisymmetric about $p/2$ (for $k =
\pi/p$), respectively.  These equations yield $s + 1$ of the $2s+1$
eigenvalues of $\mathcal{H}'$.

The remaining $s$ solutions are antisymmetric about $m=0$.  For these,
let $b_0 = 0$ and $b_m= \psi_m - \psi_{-m}= 2\psi_m$.  Then
\begin{align}
  \mathcal{V}_{m-1,k'}\, b_{m-1}
  + 2t \cos(m Q+k') \,b_m
  + \mathcal{V}_{mk'}\, b_{m+1} &= E^{- \pm}\, b_m\;\;\;
  \mathrm{for}\;\; 1\leq m < s \label{e26} \\
  \mathcal{V}_{s-1,k'}\, b_{s-1}
  + 2t \cos(s Q+k^{\prime}) \,b_s
  \pm \mathcal{V}_{sk'}\, b_s &= E^{- \pm} b_s,
  \label{e28}
\end{align}
\end{widetext}
where the $\pm$ signs again denote solutions that are symmetric about
$p/2$ (for $k = \pi$) and antisymmetric about $p/2$ (for $k = 0$).

For $k = k' = 0$, let us enumerate the eigenvalues of
(\ref{e22})--(\ref{e28}) by an index $\mu$, in order of increasing
energy.  At this point, the eigenvalues of $\mathcal{H}'$ correspond
to $E_\mu^{++}$ and $E_\mu^{--}$, and the largest eigenvalue is
$E_{s+1}^{++}$.  By inspecting the structure of the tridiagonal matrix
equations (\ref{e22})--(\ref{e25}) and (\ref{e26})--(\ref{e28}), one
can derive the relations \cite{thouless1983}
\begin{align}
  E_\mu^{--}, E_\mu^{++} \;<\; E_\mu^{-+} &\;<\; E_{\mu+1}^{++},
  \label{E_inequality} \\
  \sum_{\mu=1}^s (E_\mu^{-+}-E_\mu^{--}) &= 2 \mathcal{V}_{s0} \\
  \sum_{\mu=1}^s \left(E_\mu^{-+}-E_\mu^{++}\right) &= E_{s+1}^{++}-2t.
  \label{E_equality}
\end{align}
The bandgaps at $k = k' = 0$ lie between $E_\mu^{++}$ and
$E_\mu^{--}$, so using the above results we can derive the inequality
\begin{align}
  \begin{aligned}
    \sum_{\mu=1}^s &\left\vert E_\mu^{++} - E_\mu^{--} \right\vert \\
    &\leq
    \sum_{\mu=1}^s \Big(\left\vert E_\mu^{++}-E_\mu^{-+}\right\vert
    + \left\vert E_\mu^{-+}-E_\mu^{--}\right\vert \Big) \\
    &= E_{s+1}^{++} + 2 \mathcal{V}_{s0} - 2t.
    \label{ineq1}
  \end{aligned}
\end{align}

\begin{figure}
  \centering\includegraphics[width=0.4\textwidth]{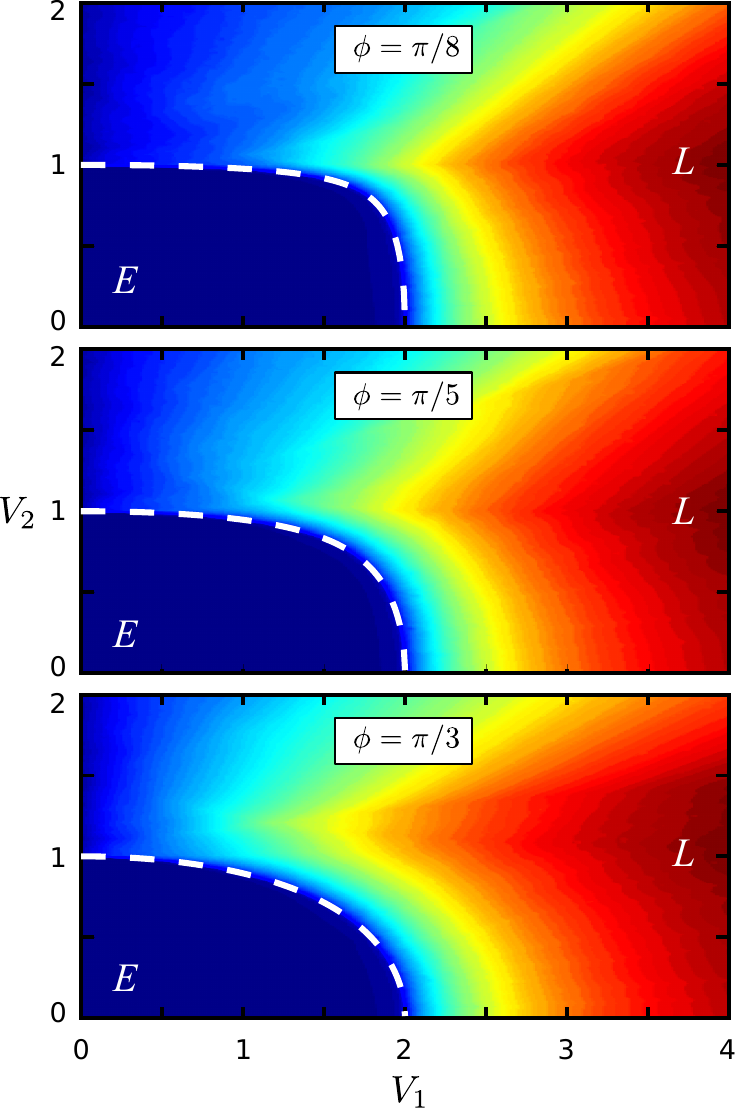}
  \caption{ Localization phase diagrams for $\phi = \pi/8$, $\phi =
    \pi/5$, and $\phi = \pi/3$, showing extended ($E$) and localized
    ($L$) phases of the generalized AAH model (\ref{AAm}).  The heat
    map shows the ground state's inverse participation ratio, with the
    largest values shown in red.  The other parameters are $t = 1$ and
    $Q = (1+ \sqrt{5})\pi$.  The dashed curves show the theoretical
    phase boundary, given by Eq.~(\ref{tran_eq2}).}
  \label{abphase}
\end{figure}

\begin{figure}
  \centering\includegraphics[width=0.45\textwidth]{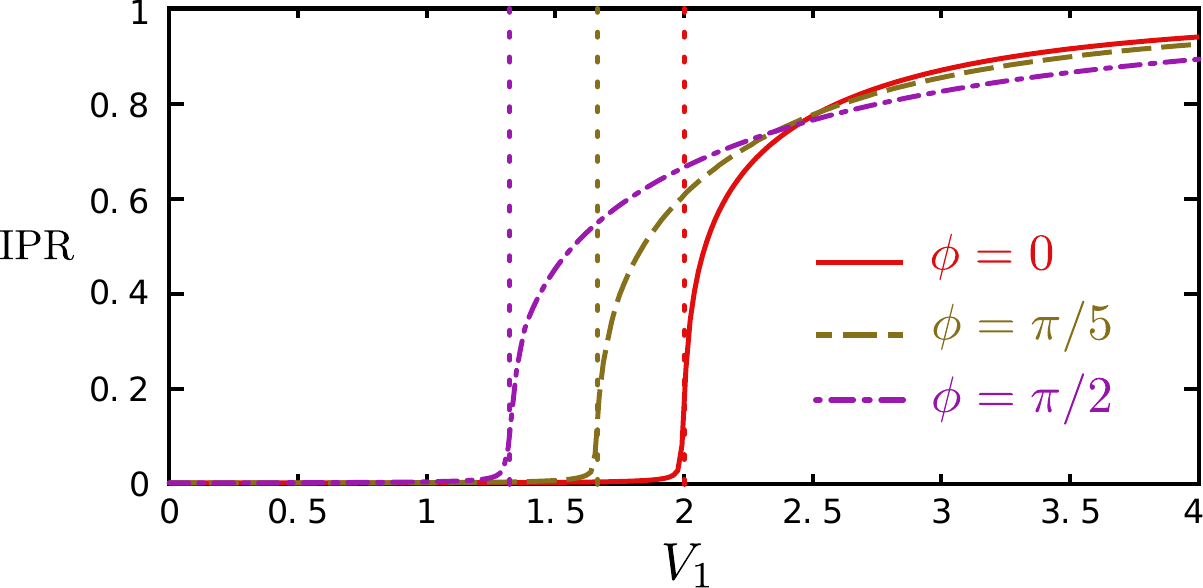}
  \caption{(Color online) Ground state inverse participation ratio
    versus $V_1$, for the generalized AAH model (\ref{AAm}),
    calculated using finite chains of length $N = 200$ with $V_2 =
    0.75$, $t = 1$, and $Q = (1+ \sqrt{5})\pi$, and three different
    values of $\phi$.  The vertical dashes show the phase boundaries
    predicted by Eq.~(\ref{tran_eq2}).}
  \label{iprv1}
\end{figure}

Next, consider $k = \pi/p$ and $k' = \pi$.  At this point, the
eigenvalues of $\mathcal{H}'$ correspond to $E^{+-}$ and $E^{-+}$.  By
inspecting Eqs.~(\ref{e26})--(\ref{e28}), we see that the $E^{-+}$
eigenvalues are the negatives of the $E^{--}$ eigenvalues which we
would have calculated at $k = k' = 0$, except using hopping amplitudes
$\mathcal{V}_{j\pi}$ instead of $\mathcal{V}_{j0}$.  The other
eigenvalues ($E^{--}$, $E^{++}$, and $E^{--}$) can all be mapped in a
similar way.  Under this mapping, the energies at the $k = \pi/p, k' =
\pi$ point will be enumerated in \textit{decreasing} order with $\mu$,
with $E_{s+1}^{+-}$ being the lowest energy.  The gaps lie between
$E_\mu^{+-}$ and $E_\mu^{-+}$, and the counterpart of (\ref{ineq1})
under the mapping is:
\begin{equation}
  \sum_{\mu=1}^s \left\vert E_\mu^{+-} - E_\mu^{-+} \right\vert
  \leq -E_{s+1}^{+-} + 2 \mathcal{V}_{s\pi} - 2t.
  \label{ineq2}
\end{equation}
The sum of the left-hand sides of (\ref{ineq1}) and (\ref{ineq2}) is
an overestimate for the sum of the bandgaps of $H'$ over the Brillouin
zone.\cite{thouless1983} The sum of the right-hand sides is
\begin{equation*}
  E^{++}_{s+1, k=k'=0}
  - E^{+-}_{s+1,k=\pi/p,k'=\pi} -4t
  + 2 \left(\mathcal{V}_{s0} + \mathcal{V}_{s\pi}\right),
\end{equation*}
where the first two terms constitute the energy difference between the
top and bottom bands.

In the incommensurate limit, therefore, the measure of the spectrum is
bounded below by
\begin{equation}
  W_{\mathrm{min}} = 4t - 2 \sum_{\pm} \left\vert\frac{V_1}{2} e^{i \phi} \pm
  V_2 \right\vert.
  \label{Wbound}
\end{equation}
From the principle that the measure vanishes at the localization
transition \cite{aubry,hofstadter}, we deduce that the transition
should occur when $W_{\mathrm{min}} = 0$; or, equivalently,
\begin{equation}
  \sum_{\pm} \sqrt{(V_1/2)^2 \pm V_1 V_2 \cos\phi + V_2^2} = 2t.
  \label{tran_eq2}
\end{equation}
In Figs.~\ref{abphase} and \ref{iprv1}, we compare this prediction for
the phase boundary to the ground state inverse participation ratios
calculated numerically from the tight-binding equation.\cite{ipr} The
results are in excellent agreement, and from this we conclude that the
eigenstates of the generalized AAH model (\ref{AAm}) are
extended---and the eigenstates of the transformed models (\ref{AA2})
and (\ref{BB2}) are localized---when
\begin{equation}
  \sum_{\pm} \sqrt{(V_1/2)^2 \pm V_1 V_2 \cos\phi + V_2^2} < 2t.
\end{equation}
This also agrees well with the localization behaviors shown in
Fig.~\ref{fig:phasediag}, for the $\phi = 0$ and $\phi = \pi/2$ cases.

\section{The period-3 generalized AAH model}
\label{Topological phase transition}

As Madsen \textit{et al.}~have recently emphasized, there is no
essential difference in the way quasicrystals and ordinary crystalline
lattices are topologically classified.\cite{madsen} In order to
understand the topological properties of the generalized AAH model
(\ref{AAm}), it is sufficient to study rational values of $Q/2\pi$.
This is equivalent to taking a ``magnetic unit cell'' of the 2D
quantum Hall lattice. \cite{TKNN}

In most of our numerical examples, we have set $Q/2\pi$ to the golden
ratio $(1+\sqrt{5})/2 = 1.61803\dots$, which is a conventional choice
for 1D quasicrystals.  To understand the resulting topological
properties, however, we have found it convenient to study a period-3
lattice with $Q/2\pi = 5/3 = 1.666\dots$~(a truncation of the golden
ratio's continued fraction).  As shown in Fig.~\ref{fig:projected}, it
accounts well for the two complete bandgaps observed in the
quasicrystal, as well as the existence of topological edge states.
The period-3 model has several interesting properties, which are
summarized in this Appendix.

Consider an infinite chain with $k = 0$.  Bloch wavefunctions satisfy
$\psi_{n+3}=e^{i K} \psi_n$, where $K \in [0,2\pi)$ denotes the
  quasi-momentum along the 1D chain.  Eq.~(\ref{AAm}) reduces to a
  eigenvalue problem $\mathcal{H}(\phi,K)\psi = E\psi$, where $\psi
  \equiv [\psi_{n-1}, \psi_{n}, \psi_{n+1}]^T$, and
\begin{equation}
  \mathcal{H}(\phi,K) =
  \begin{bmatrix}
  V_1 \cos(-Q+\phi) & \alpha & \beta\, e^{-iK}\\
  \alpha & V_1 \cos(\phi) & \alpha \\
  \beta\, e^{iK} & \alpha & V_1\cos(Q+\phi)
  \end{bmatrix},
\label{hami}
\end{equation}
where
\begin{align}
  \begin{aligned}
    \alpha &\equiv t+V_2\cos(Q/2) \\
    \beta &\equiv t+V_2\cos(3Q/2). \nonumber
  \end{aligned}
\end{align}
This Hamiltonian depends parametrically on $K$ and $\phi$.  Although
$\phi$ lacks a straightforward interpretation as a quasi-momentum, we
can nonetheless treat the parameter space spanned by $K,\phi \in
\left[0, \ 2\pi \right] $ as an abstract ``Brillouin zone''.  The
Hamiltonian obeys the symmetries
\begin{equation}
  \mathcal{H}(\phi,K) \,=\, \mathcal{H}(\phi,-K)^* \,=\, \Gamma
  \mathcal{H}(-\phi,-K) \Gamma,
  \label{symmetries}
\end{equation}
where
\begin{equation}
  \Gamma = \begin{bmatrix}0&0&1\\0&1&0\\1&0&0
  \end{bmatrix}.
\end{equation}

We can diagonalize $\mathcal{H}(\phi,K)$ for many discrete values of
$K \in [0,2\pi)$ to produce a ``projected bandstructure'' of $E$
  versus $\phi$.  Alternatively, for a finite chain of length $N \gg
  1$, the energy levels can be obtained directly from Eq.~(\ref{AAm}),
  which yields bandstructures like those shown in
  Fig.~\ref{fig:projected}.  These finite-system bandstructures,
  unlike the ones obtained from diagonalizing $\mathcal{H}(\phi,K)$,
  can contain dispersion curves corresponding to topological boundary
  states.  For fixed $V_2$, we observe that when $V_1$ is sufficiently
  small, the bandstructures appear to be topologically trivial,
  i.e.~there are no boundary states in either bandgap, as shown in
  Fig.~\ref{fig:projected}(a).  For large $V_1$, the bandstructure is
  topologically non-trivial, as in Fig.~\ref{fig:projected}(c).


\begin{figure}
  \centering
  \includegraphics[width=0.47\textwidth]{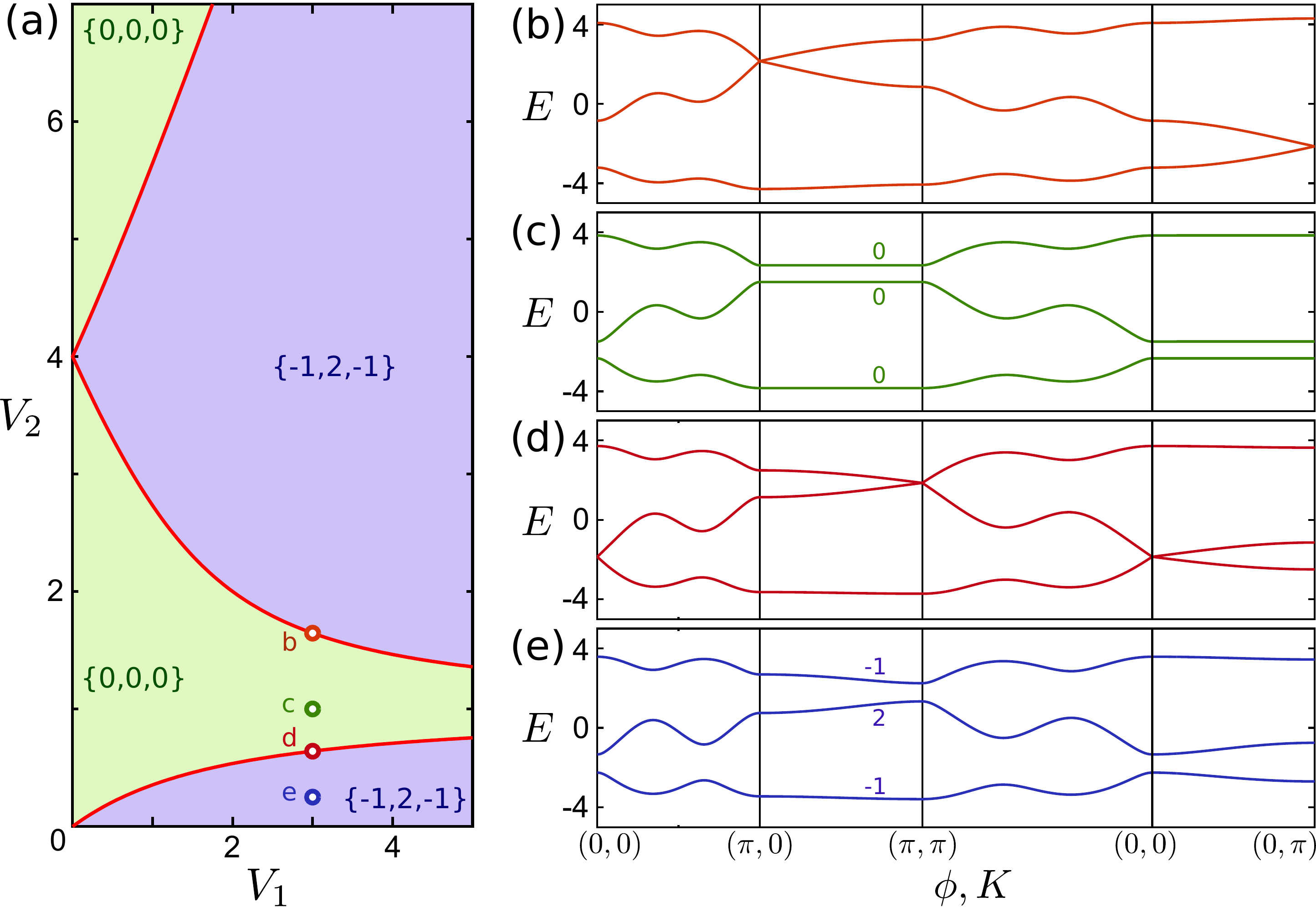}
  \caption{(Color online) (a) Topological phase diagram of the
    period-3 AAH model.  The phase boundaries are given by
    Eqs.~(\ref{trans})--(\ref{trans2}), and labels indicate the bands'
    Chern numbers.  (b)--(e) Band diagrams at the values of $V_1$ and
    $V_2$ indicated in (a): (b) $V_1 = 3, V_2 = 1.6458$; (c) $V_1 = 3,
    V_2 = 1$; (d) $V_1 = 3, V_2 = 0.64110$; (d) $V_1 = 3, V_2 = 0.25$.
    The gapped bands are labeled by their Chern numbers.  The other
    parameters are $t = 1$, $Q = 10\pi/3$, and $k = 0$.}
  \label{fig4}
\end{figure}

In order to understand the topological phase diagram quantitatively,
we look for bandgap closings in the bulk Hamiltonian
$\mathcal{H}(\phi,K)$.  The symmetry relations (\ref{symmetries})
indicate that we can focus on the high-symmetry points in the
Brillouin zone: $(0,0)$, $(\pi, \pi)$, $(\pi, 0)$, and $(0,\pi)$.  At
these points, the Hamiltonian has another important property, which
can be seen from the matrix structure of Eq.~(\ref{hami}): the
eigenvalues of $\mathcal{H}(0,0)$ are the negatives of the eigenvalues
of $\mathcal{H}(\pi,\pi)$, and the eigenvalues of $\mathcal{H}(0,\pi)$
are the negatives of the eigenvalues of $\mathcal{H}(\pi,0)$.  As a
result, bandgap closings always occur in pairs, at different points in
the Brillouin zone.  But, unlike familiar cases such as graphene,
these simultaneous pairwise closings occur at different energies, in
different bandgaps.

By diagonalizing $\mathcal{H}(0,0)$, we obtain the eigenvalues
 \begin{align}
   \begin{aligned}
     \mathcal{E}_0&= -t +V_1 \cos(Q)- V_2 \cos(3Q/2) \label{eqE0} \\
     \mathcal{E}_\pm&=\frac{Z \pm W}{2},
   \end{aligned}
 \end{align}
where
\begin{align}
  Z & =t+V_1+V_1\cos(Q) +V_2 \cos(3Q/2) \\
  W^2 &= Z^2 + 8 \Big[t+V_2\cos(Q/2)\Big]^2 \nonumber\\
  &\;\;\qquad -4V_1\Big[t + V_1 \cos(Q) + V_2 \cos(3Q/2)\Big].
  \label{eqW}
\end{align}
We now set $Q=10\pi/3$ for specificity.  The bandgaps at $(0,0)$ and
$(\pi,\pi)$ close when $\mathcal{E}_0 = \mathcal{E}_\pm$; this yields
the phase boundaries
\begin{equation}
  V_2 = 2t + V_1 \pm \sqrt{(t+V_1)^2+3t^2},
  \label{trans}
\end{equation}
or, equivalently,
\begin{equation}
  V_1 = \frac{V_2}{2} \,\left( \frac{4t - V_2}{t - V_2}\right).
\end{equation}

Next, consider $\mathcal{H}(0,\pi)$.  Following a similar procedure,
we find that the bandgaps at $(0,\pi)$ and $(\pi,0)$ close when
\begin{equation}
  V_2 = 2t -V_1 + \sqrt{(t-V_1)^2 + 3t^2}.
  \label{trans2}
\end{equation}
(The $-$ solution is discarded since it gives negative $V_2$; by
convention, the modulation amplitudes are positive.)

As shown in Fig.~\ref{fig4}(a), Eqs.~(\ref{trans}) and (\ref{trans2})
partition the $\{V_1,V_2\}$ phase space into four distinct gapped
phases.  We can characterize the topology of the gapped bandstructures
by calculating each band's Chern number,\cite{TKNN}
\begin{align}
  C_n &= \frac{1}{2\pi i}\int\!\!\int_{BZ} d\phi\; dK\, \left(\frac{\partial A_{nn}^K}{\partial \phi}- \frac{\partial A_{nn}^\phi}{\partial K}\right),
  \label{Chern} \\
  \vec{A}_{n n'} &= \Big\langle n, \phi, K\Big| \vec{\nabla}_{\phi,K}
  \Big| n', \phi, K \Big\rangle.
  \label{Berry}
\end{align}
Here, $| n, \phi, K\rangle$ denotes the Bloch eigenstate in the $n$th
band at parameter values $(\phi, K)$.  As indicated in
Fig.~\ref{fig4}(a), the topologically trivial bandstructures have
Chern numbers $\{0,0,0\}$, while the topologically non-trivial
bandstructures have Chern numbers $\{-1,2,-1\}$.

The bottom-most phase boundary in Fig.~\ref{fig4}(a), which
corresponds to the $-$ solution of Eq.~(\ref{trans}), cuts across the
localization phase boundary of the incommensurate generalized AAH
model (see Figs.~\ref{fig:phasediag} and \ref{abphase}).  When the
incommensurate model is in the extended phase, or not too deep into
the localized phase, we find that its topological behavior closely
matches the behavior of the period-3 model.  Specifically, its
bandstructure contains two primary complete bandgaps, which occur at
energies similar to the bandgaps of the period-3 model and have
similar topological transitions, as shown in Figs.~\ref{fig:projected}
and \ref{fig:endstates}.  This correspondence appears to break down,
however, deep in the localized phase.  Future studies will seek a
better understanding of the incommensurate model's topological
properties within the strongly localized regime.

\end{document}